\documentclass[10pt]{article}

\usepackage[T1]{fontenc}
\usepackage[utf8]{inputenc}
\usepackage{graphicx, verbatim, url, amssymb, amsfonts, amsthm, latexsym, hyperref, caption, subcaption, rotating}
\usepackage{authblk}
\usepackage{booktabs}
\usepackage[fleqn]{amsmath}
\usepackage{multirow}
\usepackage{csquotes}

\title{Wikipedia Citations:\\ A comprehensive dataset of citations with identifiers extracted from English Wikipedia}

\author[1]{Harshdeep Singh}
\author[1]{Robert West}
\author[2]{Giovanni Colavizza\thanks{Corresponding author, g.colavizza@uva.nl}}

\affil[1]{Data Science Laboratory, EPFL}
\affil[2]{Institute for Logic, Language and Computation, University of Amsterdam}

\begin{document}
	
\maketitle

\begin{abstract}
	Wikipedia's contents are based on reliable and published sources. To this date, relatively little is known about what sources Wikipedia relies on, in part because extracting citations and identifying cited sources is challenging. To close this gap, we release \texttt{Wikipedia Citations}, a comprehensive dataset of citations extracted from Wikipedia. A total of 29.3M citations were extracted from 6.1M English Wikipedia articles as of May 2020, and classified as being to books, journal articles or Web contents. We were thus able to extract 4.0M citations to scholarly publications with known identifiers --- including DOI, PMC, PMID, and ISBN --- and further equip an extra 261K citations with DOIs from Crossref. As a result, we find that 6.7\% of Wikipedia articles cite at least one journal article with an associated DOI, and that Wikipedia cites just 2\% of all articles with a DOI currently indexed in the Web of Science. We release our code to allow the community to extend upon our work and update the dataset in the future.  
\end{abstract}


\section{Introduction}

\begin{displayquote}
\textit{``Citations have several important purposes: to uphold intellectual honesty (or avoiding plagiarism), to attribute prior or unoriginal work and ideas to the correct sources, to allow the reader to determine independently whether the referenced material supports the author's argument in the claimed way, and to help the reader gauge the strength and validity of the material the author has used.''}\footnote{\url{https://en.wikipedia.org/wiki/Citation} [accessed 2020-01-03].}
\end{displayquote}

\noindent Wikipedia plays a fundamental role as a source of factual information on the Web: it is widely used by individual users as well as third-party services, such as search engines and knowledge bases~\cite{lehmann_dbpedialarge-scale_2015,mcmahon_substantial_2017}.\footnote{\url{https://en.wikipedia.org/wiki/Wikipedia:Statistics} [accessed 2020-01-03].} Most importantly, Wikipedia is often perceived as a source of reliable information~\cite{mesgari_sum_2015}. The reliability of Wikipedia's contents has been debated, since anyone can edit them~\cite{okoli_peoples_2012,mesgari_sum_2015}. Nevertheless, the confidence that users and third-party services place on Wikipedia appears to be justified: Wikipedia's contents are of general high-quality and up-to-date, as shown by several studies over time~\cite{priedhorsky_creating_2007,keegan_hot_2011,okoli_peoples_2012,geiger_when_2013,kumar_disinformation_2016,piscopo_what_2019,Colavizza_2020}.

To reach this goal, Wikipedia's verifiability policy mandates that ``people using the encyclopedia can check that the information comes from a reliable source.'' A reliable source is defined, in turn, as a secondary and published, ideally scholarly one.\footnote{See respectively \url{https://en.wikipedia.org/wiki/Wikipedia:Verifiability} and \url{https://en.wikipedia.org/wiki/Wikipedia:Reliable_sources} [accessed 2020-01-03].} Despite the community's best efforts to add all the needed citations, the majority of articles in Wikipedia might still contain unverified claims, in particular lower-quality ones~\cite{damasevicius_analysis_2017}. The citation practices of editors might also be at times not systematic~\cite{chen_citation_2012,forte_information_2018}. As a consequence, the efforts to expand and improve Wikipedia's verifiability through citations to reliable sources are increasing~\cite{fetahu_finding_2016,redi_citation_2019}.

A crucial question to ask in order to improve Wikipedia's verifiability standards, as well as to better understand its dominant role as a source of information, is the following: \textit{What sources are cited in Wikipedia?}

A high portion of citations to sources in Wikipedia refer to scientific or scholarly literature~\cite{blomqvist_scholia_2017}, as Wikipedia is instrumental in providing access to scientific information and in fostering the public understanding of science~\cite{laurent_seeking_2009,heilman_wikipedia_2011,damasevicius_analysis_2017,shafee_evolution_2017,maggio_wikipedia_2019,torres-salinas_mapping_2019,maggio_reader_2020,smith_situating_2020}. Citations in Wikipedia are also useful for users browsing low-quality or underdeveloped articles, as they allow them to look for information outside of the platform~\cite{piccardi_quantifying_2020}. The literature cited in Wikipedia has been found to positively correlate with a journal's popularity, journal impact factor and open access policy~\cite{nielsen_scientific_2007,teplitskiy_amplifying_2017,arroyo-machado_science_2020}. Being cited in Wikipedia can also be considered as an `altmetric' indicator of impact in itself~\cite{sugimoto_scholarly_2017,kousha_are_2017}. A clear influence of Wikipedia on scientific research has in turn been found~\cite{thompson_science_2018}, despite a general lack of reciprocity in acknowledging it as a source of information from the scientific literature~\cite{jemielniak_bridging_2016,tomaszewski_study_2016}. Nevertheless, the evidence on what scientific and scholarly literature is cited in Wikipedia is quite slim. Early studies point to a relative low coverage, indicating that between 1\% and 5\% of all published journal articles are cited in Wikipedia~\cite{priem_altmetrics_2012,shuai_comparative_2013,zahedi_how_2014,pooladian_methodological_2017}. These studies possess a number of limitations: they consider a by-now dated version of Wikipedia, they use proprietary citation databases with limited coverage, or they only consider specific publishers (PLoS) and academic communities (computer science, library and information science). More recently, a novel dataset has been released containing the edit history of all references in English Wikipedia, until June 2019~\cite{zagovora_i_2020}. While the authors found a persistent increase of references equipped with some form of document identifier over time, they underline how relying on references with document identifiers is still not sufficient to capture all relevant publications cited from Wikipedia.

Answering the question of what exactly is cited in Wikipedia is challenging for a variety of reasons. First of all, editorial practices are not uniform, in particular across different language versions of Wikipedia: citations are often given using citation templates somewhat liberally,\footnote{\url{https://en.wikipedia.org/wiki/Wikipedia:Citation_templates} [accessed 2020-01-03].} making it difficult to detect citations to the same source. Secondly, while some citations contain stable identifiers (e.g., DOIs), others do not. A recent study found that 4.42\% Wikipedia articles contain at least one citation with a DOI~\cite{maggio_wikipedia_2019}: a low fraction which might indicate that we are missing a non-negligible share of citations without identifiers. This is a significant limitation since existing databases, such as Altmetrics, do provide Wikipedia citation metrics relying exclusively on citations with identifiers.\footnote{\url{https://help.altmetric.com/support/solutions/articles/6000060980-how-does-altmetric-track-mentions-on-wikipedia} [accessed 2020-01-03]. Identifiers considered by Altmetrics currently include: DOI, URI from a domain white list, PMID, PMC ID, arXiv ID.} This in turn limits the scope of results relying on these data.

Our goal is to overcome these two challenges and expand upon previous work~\cite{halfaker_mansurov_redi_taraborelli_2018}, by providing a dataset of \textit{all} citations from English Wikipedia, equipped with identifiers and including the code necessary to replicate and improve upon our work. The resulting dataset is available on Zenodo~\cite{dataset}, while an accompanying repository contains code and further documentation.\footnote{\url{https://github.com/Harshdeep1996/cite-classifications-wiki/releases/tag/0.2}.} By releasing a codebase which allows to extract citations directly from Wikipedia data, we aim to address the following limitations found in previous work: the focus on specific publishers or scientific communities, the use of proprietary databases, and the lack of means to replicate and update results. Given how dynamic Wikipedia is, we deem of importance to release a codebase to keep the \texttt{Wikipedia Citations} dataset up to date for future re-use. 

This article is organized as follows. We start by describing our pipeline focusing on its three main steps: 1) citation template harmonization --- to structure every citations in Wikipedia using the same schema; 2) citation classification --- to find citations to books and journal articles; and 3) citation identifier look-up --- to find identifiers such as DOIs. We subsequently evaluate our results, provide a description of the published dataset, and conclude by highlighting some possible uses of the dataset as well as ideas to improve it further.

\section{Methodology}

\begin{figure*}
	\centering
	\includegraphics[width=\linewidth]{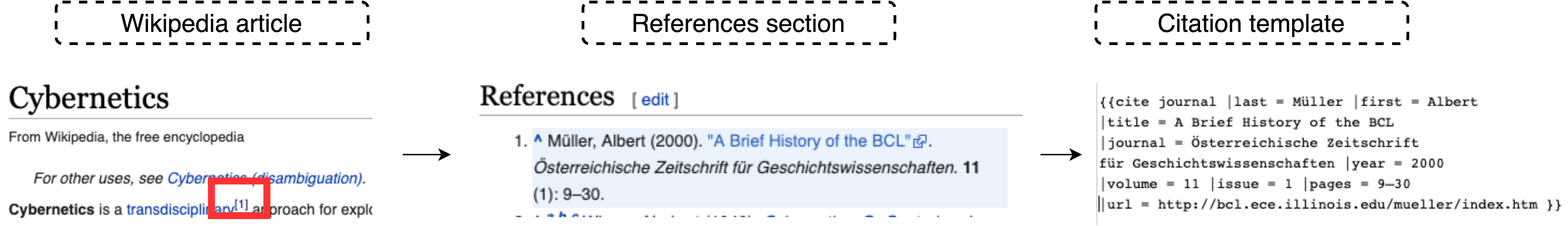}
	\caption{Example of citations in Wikipedia.}
	\label{fig:reference_example}
\end{figure*}

\noindent We start by briefly introducing Wikipedia-specific terminology:

\begin{itemize}
	\item \textit{Wikicode}: The markup language used to write Wikipedia pages; also known as Wikitext or Wiki markup.
	\item \textit{Template}: A page that is embedded into other pages to allow for the repetition of information, following a certain Wikicode format.\footnote{\url{https://en.wikipedia.org/wiki/Help:Template} [accessed 2020-01-03].} Citation templates are specifically defined to embed citations.
	\item \textit{Citation}: A citation is an abbreviated alphanumeric expression, embedded in Wikicode following a citation template, as shown in Figure \ref{fig:reference_example}; it usually denotes an entry in the \textit{References} section of a Wikipedia page, but can be used anywhere on a page too (e.g., \textit{Notes}, \textit{Further work}).
	
\end{itemize}

\subsection{Overview}
Our process can be broken down into the following steps, as illustrated in Figure \ref{fig:overview_pipeline}:

\begin{figure*}
	\centering
	\includegraphics[width=\linewidth]{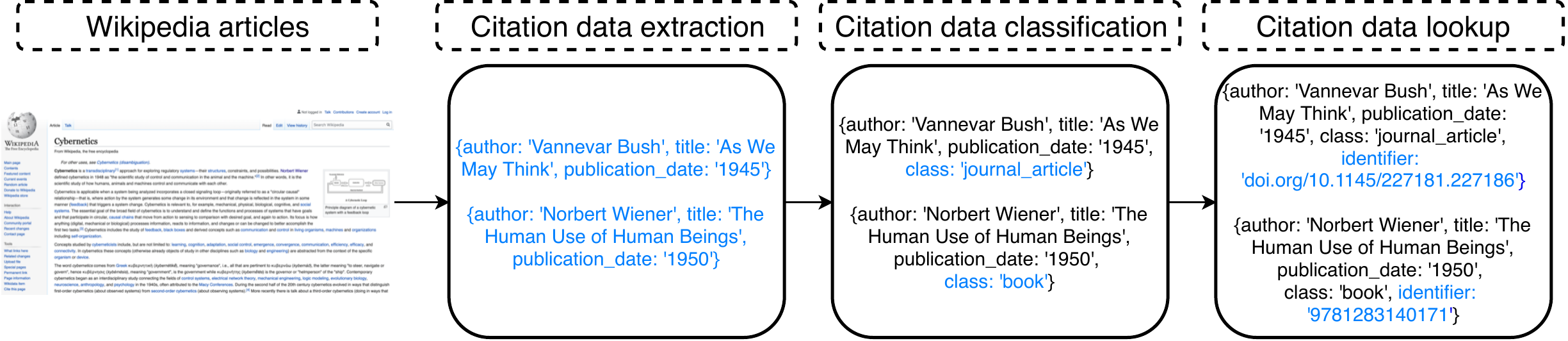}
	\caption{Overview of the citation data extraction pipeline. We highlight in blue/grey the outputs at every stage. These examples are illustrative simplifications from the actual dataset.}
	\label{fig:overview_pipeline}
\end{figure*}

\begin{enumerate}
	\item \textit{Citation data extraction}: A Wikipedia dump is used to extract citations from all pages and considering various citation templates. The extracted citations are then mapped to a uniform set of key-value pairings.
	\item \textit{Citation data classification}: A classifier is trained to distinguish between citations to journal articles, books, or other Web content. The classifier is trained using a subset of citations already equipped known identifiers or URLs, allowing to label them beforehand. All the remaining citations are then classified.
	\item \textit{Citation data lookup}: All newly found citations to journal articles are labeled with identifiers (DOIs) using the Crossref API.
\end{enumerate}

\subsection{Citation data extraction}

The citation data extraction pipeline is in turn divided into two steps, which are repeated for every Wikipedia article: 1) \textit{extraction} of all sentences which contain text in Wikicode format, and \textit{filtering} of sentences using the citation template Wikicode; 2) \textit{mapping} of extracted citations to the uniform template and creation of a tabular dataset. An example of Wikicode citations, extracted during step 1, is given in Table \ref{tab:differentkey}. The same citations after mapping to a uniform template are given in Table \ref{tab:differentdict__}.

\subsubsection{Extraction and filtering}
We used the English Wikipedia XML dump from May 2020 and scraped it to get the content of each article/page. The number of unique pages is 6,069,685 after removing redirects since they do not have any citations of their own.
\par
Since we are restricting ourselves to citations which are given in Wikicode format, we used the \texttt{mwparserfromhell} parser,\footnote{\url{https://github.com/earwig/mwparserfromhell} [version 0.6].} which given as input a Wikipedia page, it returns all text which is written in Wikicode format. Citations are generally present inside \texttt{<ref>} tags or between double curly brackets \texttt{\{\{}, as shown in Table \ref{tab:differentkey}. When multiple citations to the same source are given in a page, we only consider the first one. The number of extracted citations is 29,276,667.

\subsubsection{Mapping}
Citation templates can vary, and different templates can be used to refer to the same source in different pages. Therefore, we mapped all citations to the same uniform template. For this step, we used the \texttt{wikiciteparser} parser.\footnote{\url{https://github.com/dissemin/wikiciteparser} [version 0.1.1].} This parser is written in \textit{Lua} and it can be imported into \textit{Python} using the \texttt{lupa} library.\footnote{\url{https://pypi.org/project/lupa}.} The uniform template we use comprises 29 different keys. Initially, the \texttt{wikiciteparser} parser only supported 17 citation templates, thus we added support for an additional 18 of the most frequently used templates. More details on the uniform template keys and the extra templates we implemented can be found in the accompanying repository.
\par
The resulting uniform key-value dataset can easily be transformed in tabular form for further processing. In particular, this first step allowed us to construct a dataset of citations with identifiers containing approximately 3.928 million citations. These identifiers --- including DOI, PMC, PMID and ISBN --- allowed us to use such citations as training data for the classifier.

\subsection{Citation data classification}

After having extracted all citations and mapped them to a uniform template, we proceed to train a classifier to distinguish among three categories of cited sources: \textit{journal articles, books and Web content}. Our primary focus are journal articles, as those cover most citations to scientific sources. We describe here our approach to label a golden dataset to use for training, the features we use for the classifier, and the classification model.

\subsubsection{Labeling}
We labelled available citations as follows:
\begin{itemize}
	\item Every citation with a PMC or PMID was labeled as a \textit{journal article}.
	\item Every citation with a PMC, PMID or DOI and using the citation template for journals and conferences, was labeled as a \textit{journal article}.
	\item Every citation which had an ISBN was labelled as a \textit{book}.
	\item All citations with their URL top level domain belonging to the following: \textit{nytimes, bbc, washingtonpost, cnn, theguardian, huffingtonpost, indiatimes}, were labeled as \textit{Web content}.
	\item All citations with their URL top level domain belonging to the following: \textit{youtube, rollingstone, billboard, mtv, metacritic, discogs, allmusic}, were labeled as \textit{Web content}.
\end{itemize}
After labelling, we removed all identifiers and the type of citation template as features, since they were used to label the dataset. We also removed the fields: \textit{URL, work, newspaper, website}, for the same reason. The final number of data points used for training and testing the classifier is given in Table \ref{tab:train_dataset}, and was partially sampled in order to have a comparable number of journal articles, books and Web content.

\begin{table}[ht]
	\caption{Number of citations with a known class (* indicates a sampled subset).}
	\centering
	\begin{tabular}{@{}l r r p{0.37\linewidth}p{0.25\linewidth}p{0.25\linewidth}@{}}
		\hline
		Class label & Train data & Total known\\
		\hline
		Book & *951,991 & 2,103,492\\
		Web content & *1,100,000  & 3,409,042\\
		Journal article & *748,009 & 1,482,040\\
		\hline
		Total & 2,800,000 & 6,994,574 \\
		\hline
	\end{tabular}
	\label{tab:train_dataset}
\end{table}

\subsubsection{Features}
We next describe the features we used for the classification model:
\begin{itemize}
	\item \textit{Citation text}: The text of the citation, in Wikicode syntax.   
	\item \textit{Citation statement}: The text preceding a citation in a Wikipedia article, as it is known that certain statements are more likely to contain citations~\cite{redi_citation_2019}. We have used the 40 words preceding the first time a source is cited in an article.
	\item \textit{Part of Speech (POS) tags}: POS tags in citation statements could also help qualify citations~\cite{redi_citation_2019}. These were generated using the NLTK library.\footnote{\url{https://www.nltk.org} [version 3.4.1].}
	\item \textit{Citation section}: The article section a citation occurs in.
	\item \textit{Order of the citation} within the article, and \textit{total number of words of the article}.
\end{itemize}

\subsubsection{Classification model}
\begin{figure}
	\centering
	\includegraphics[width=\linewidth]{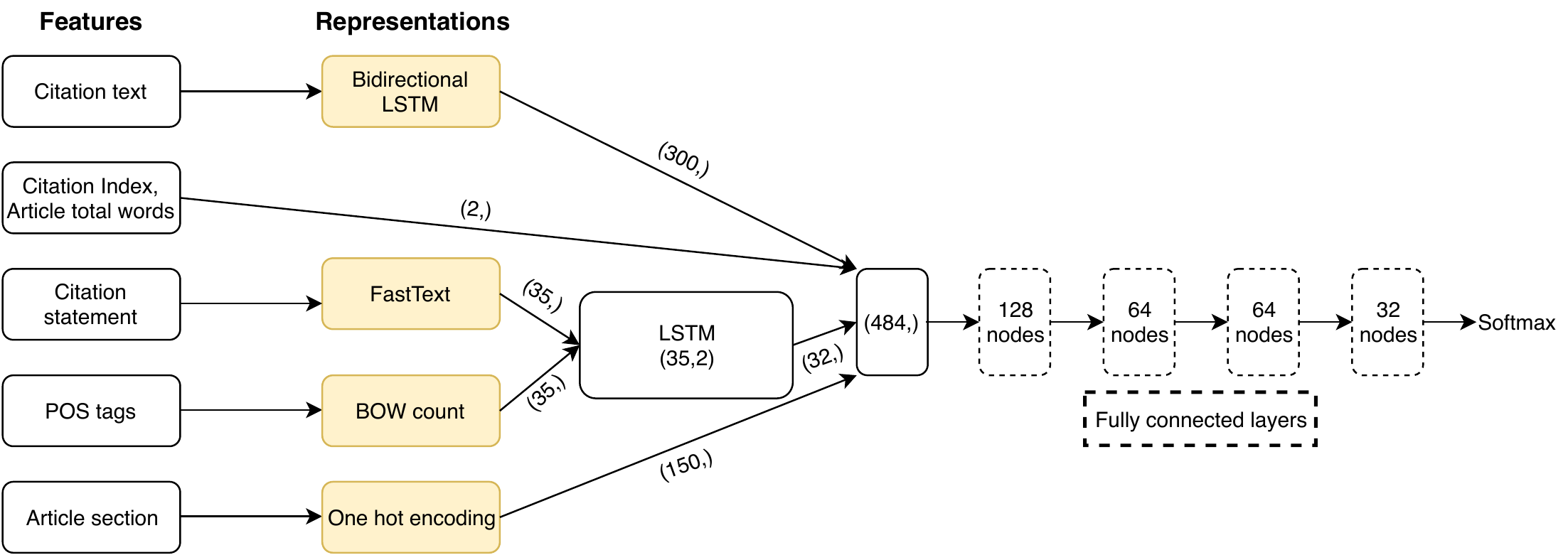}
	\caption{Citation classification model.}
	\label{fig:model_pipeline}
\end{figure}

The model which we constructed is a hybrid deep learning pipeline illustrated in Figure \ref{fig:model_pipeline}. The features were represented as follows:
\begin{itemize}
	\item \textit{Citation text}: The citation text in Wikicode syntax was fed to a character-level bidirectional LSTM \cite{schuster1997bidirectional} on the dummy task of predicting whether the citation text is to a book/journal article or other Web content. The train-test split was done using a 90-10 ratio, yielding a 98.56\% test accuracy. We used this dummy task in order to avoid the effects of vocabulary sparsity due to Wikicode syntax. The character-level embeddings are of dimension 300, we aggregated them for every citation text via summation and normalized the resulting vector to sum to one. We used character-level embeddings to deal with Wikicode syntax. The citation text embeddings were trained on the dummy task and froze afterwards.
	\item \textit{Citation statement}: The vocabulary for citation statements contains approximately 443,000 unique tokens, after the removal of tokens which appear strictly less than 5 times in the corpus. We used fastText to generate word-level embeddings for citation statements, using subword information~\cite{DBLP:journals/corr/BojanowskiGJM16}. FastText allowed us to deal with out of vocabulary words. We used the fastText model pre-trained on English Wikipedia.\footnote{\url{https://fasttext.cc/docs/en/crawl-vectors.html}.}
	\item \textit{POS tags}: The POS tags of citation statements were represented with a bag of words count vector. We were considering the top 35 tags by count frequency.
	\item \textit{Citation section}: We used a one-hot encoding for the 150 most common sections within Wikipedia articles. The \textit{order of the citation} within the article and \textit{total number of words of the article} were represented as scalars.
\end{itemize}
Once the features had been generated, citation statements and their POS tags were further fed to an LSTM of 64 dimensions to create a single representation. All the resulting feature representations were concatenated and fed into a fully connected neural network with four hidden layers, as shown in Figure \ref{fig:model_pipeline}. A final Softmax activation function was applied on the output generated by the fully connected layers, to map the output to one of the three categories of interest. We trained the model for five epochs using a train and test split of 90\% and 10\% respectively. For training, we used the Adam optimizer \cite{kingma2014adam} and a binary cross-entropy loss. The model's initial learning rate was set to $10^{-3}$, and reduced minimally to $10^{-5}$ once the accuracy metric has stopped improving.

Previous work on citation classification has been based on textual and syntactic features~\cite{dong2011ensemble} or link type analysis~\cite{xu2013using}. Different stylistic or rhetorical cues have been also used as features~\cite{di2006using}. We note that most of this previous work has focused on classifying the function or intent of a citation, rather than the typology of cited object -- that is, what a citations refers to, for example a book or journal.

\subsection{Citation data lookup}

The lookup task entails finding a permanent identifier for every citation missing one. We focused on journal articles for this final step, since they make up the bulk of citations to scientific literature found in Wikipedia for which a stable identifier can be retrieved. We used the Crossref API to get DOIs.\footnote{\url{https://www.crossref.org}.} Crossref allows to query its API 50 times per second, we used the \texttt{aiohttp} and \texttt{asyncio} libraries to process requests asynchronously. For each citation query, we get a list of possible matches in descending ordered according to a Crossref confidence score. We kept the top three results from each query response.

\section{Evaluation}

In this section we discuss the evaluation of the citation classification and lookup steps.

\subsection{Classification Evaluation}
After training the model for five epochs, we attained an accuracy of 98.32\% on the test set. The confusion matrix for each of the labels is given in Table \ref{tab:cf_random}. The model is able to distinguish among the three classes very well.

\begin{table}[ht]
	\caption{Confusion matrix for citation classification. Results are based on a 10\% held-out test set.}
	\centering
		\begin{tabular}{@{}l r r r p{0.1\linewidth}p{0.2\linewidth}p{0.15\linewidth}p{0.15\linewidth}@{}}
		\hline
		Label & Book & Article & Web\\
		\hline
		Book & 93,602 (98.32\%) & 1039  & 558 \\
		Article & 961 & 73,682 (98.50\%) & 158 \\
		Web & 1,136 & 180 & 108,684 (98.80\%) \\
		\hline
	\end{tabular}
	\label{tab:cf_random}
\end{table}

The model was then used to classify all the remaining citations from the 29.276 million dataset, that is to say approximately 22.282 million citations. Some examples of results from the classification step are given in Table \ref{tab:differentdict}.
The resulting total number of citations per class are given in Table \ref{tab:predicted_we_results}. 

\begin{table}[ht]
\caption{Number of newly-classified citations per class.}
\centering
		\begin{tabular}{@{}l r r r p{0.2\linewidth}p{0.2\linewidth}p{0.2\linewidth}p{0.2\linewidth}@{}}
\hline
Label & New & Previously known & Total\\ 
\hline
Journal article & 947,233 & 1,482,040 & 2,429,273 \\
Book & 3,243,364 & 2,103,492 & 5,346,856 \\
Web content & 18,091,496 & 3,409,042 & 21,500,538 \\
\hline
Total & 22,282,093 & 6,994,574 & 29,276,667 \\ 
\hline
\end{tabular}
\label{tab:predicted_we_results}
\end{table}

\subsection{Crossref Evaluation}
For the lookup, we evaluated the response of the Crossref API in order to assess how to select results from it. We tested the API using 10,000 random citations with DOI identifiers and containing 9764 unique title-author pairs. We split this subset into a 80-20 split, tried out different heuristics on 80\% of the data points and tested the best one on the remaining 20\%. Table \ref{tab:Crossref_eval} shows the results for different heuristics, which confirms that the simple heuristic of picking the first result from Crossref works well. 

\begin{figure}
	\centering
	\begin{subfigure}{\textwidth}
	    \centering
		\includegraphics[width=0.9\linewidth]{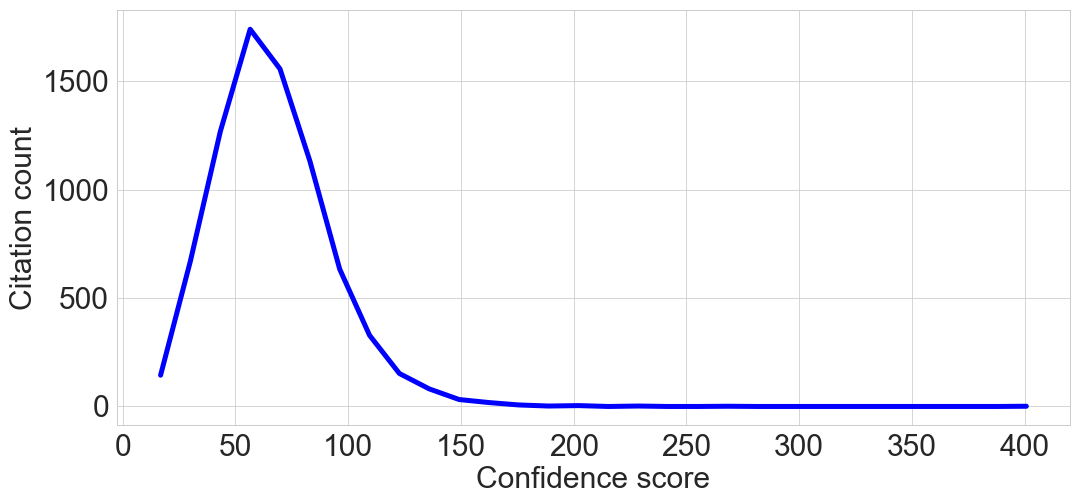}
		\caption{Histogram of the Crossref API confidence scores over the validation set of the first result extracted from the lookup.}
		\label{fig:sub1_crossref}
	\end{subfigure}%
	\\
	\begin{subfigure}{\textwidth}
	    \centering
		\includegraphics[width=0.9\linewidth]{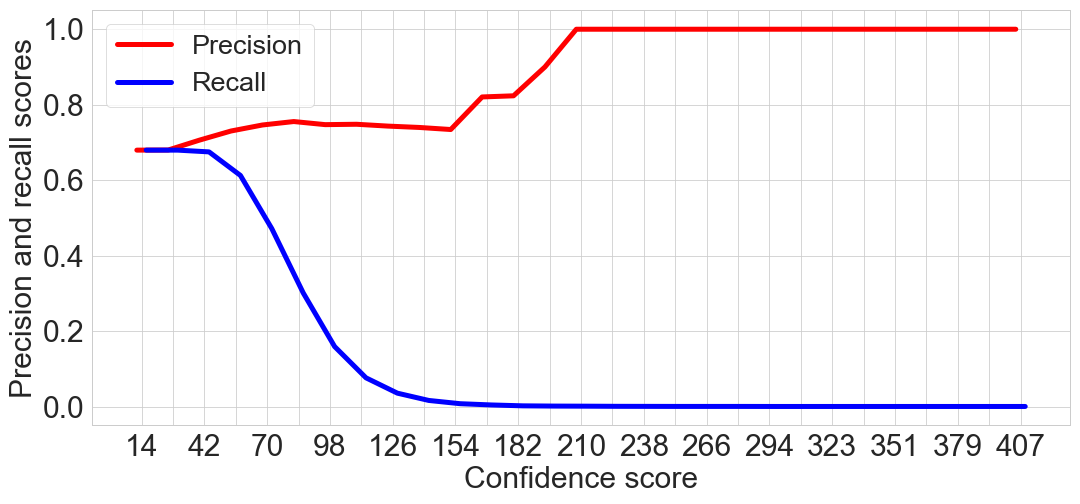}
		\caption{Precision and recall for different Crossref API confidence score thresholds where the x-axis represents the scores returned by the Crossref API.}
		\label{fig:sub2_crossref}
	\end{subfigure}
	\caption{Evaluation of the Crossref API scores.}
\label{fig:all_pubs}
\end{figure}

This still leaves open the question of what Crossref confidence score to use. We picked the threshold for the confidence score to be 34.997 which gave us a precision of 70\% and a recall of 67.55\% to reach a balance between the two in the evaluation (Figure \ref{fig:all_pubs}).

\begin{table}[ht]
	\caption{Results for each heuristic tested on 80\% of the subset.}
	\centering
	\begin{tabular}{@{}l r r r p{0.17\linewidth}p{0.13\linewidth}p{0.15\linewidth}p{0.15\linewidth}@{}}
		\hline
		Heuristic & Matched & Not matched & Invalid request\\
		\hline
		1st result & 5258 & 2510 & 43 \\
		2nd result & 345 & 7407 & 59  \\
		3rd result & 96 & 7647 & 67\\
		\hline
	\end{tabular}
	\label{tab:Crossref_eval}
\end{table}

We finally tested the threshold using the 1953 held-out examples, out of which 1246 examples had the correct identifier with the first heuristic (out of 1297) and the threshold, 646 examples gave a different result out of which 521 are over the threshold and only 10 requests were invalid for the API. Hence, the first metadata result is the best result from the Crossref API. 

The lookup process was performed by extracting the title and the first author (if available) for all the potential journal articles and was queried against the CrossRef API to get the metadata. The top 3 results from the metadata were taken into account if they existed, and their DOIs and confidence scores were extracted. 260,752 citations were equipped with DOIs using the lookup step and 153,879 unique DOIs were found relating to each of these citations (selecting the DOI with highest Crossref score).

\section{Dataset}
\noindent The resulting \texttt{Wikipedia Citations} dataset is composed of 3 parts:

\begin{enumerate}
	\item The main dataset of 29.276 million citations from 35 different citation templates, out of which 3.928 million citations already contained identifiers (Table \ref{tab:identifier_per_citation}), and 260,752 out of 947,233 newly-classified citations to journal articles were equipped with DOIs from Crossref.
	\item An example subset with the features for the classifier.
	\item Citations classified as journal and their corresponding metadata/identifier extracted from Crossref to make the dataset more complete.
\end{enumerate}

\subsection{Descriptive analysis}
We start by comparing our dataset with previous work, which focused on citations with identifiers~\cite{halfaker_mansurov_redi_taraborelli_2018}. The total number of citations per identifier type is found to be similar (Table \ref{tab:differencedataset}). Minor discrepancies are likely due to the fact that we do not consider here all the edit history of every Wikipedia page, therefore missing changes between revisions, and that we consider a more recent dump. The total number of distinct identifiers across all Wikipedia, both previously known and newly-found, are given in Table \ref{tab:unique_identifiers}. Considering that in the Web of Science (WoS)~\cite{birkle_web_2020} at the time there were 34,640,325 unique DOIs (version of June 2020; we only consider the typologies of `article', `review', `letter' and `proceedings paper'), Wikipedia is citing a volume of unique DOIs (1,157,571) corresponding to 3.3\% of this total. Yet by doing an exact matching between Wikipedia DOIs and Web of Science DOIs, we can find 710,913 identifiers which are in common, or just 2\% of the Web of Science total. This also entails that approximately 61\% of unique DOIs in Wikipedia are indexed in the Web of Science. This result is in line with previous findings~\cite{priem_altmetrics_2012,shuai_comparative_2013,zahedi_how_2014,pooladian_methodological_2017}. The proportion of cited articles might seem low when compared to all of science, yet it is worth considering that an editorial selection takes place: articles cited from Wikipedia are typically highly cited and published in visible journals~\cite{nielsen_scientific_2007,teplitskiy_amplifying_2017,arroyo-machado_science_2020,Colavizza_2020}. All in all, the relatively low fraction of scientific articles cited from Wikipedia over the total available, does not \textit{per se} entail a lack of coverage or quality in its contents: more work is needed to assess whether this might be the case.

We next consider the Web of Science subject categories for these 710,913 articles. We list the top-30 subject categories in Table~\ref{tab:wos_sc}, by number of distinct articles cited from Wikipedia. This ranking is dominated by Biochemistry \& Molecular Biology (more than 11\% of the articles) and Multidisciplinary Sciences (7\%). The latter category accounts for mega journals such as Nature, Science and PNAS. In general, the life sciences and biomedicine dominate. The top social science is Economics (1\%) and the top humanities discipline is History (0.9\%). To be sure, these results should be taken with caution, in particular when considering the arts, humanities and social sciences. In this respect, the coverage of the Web of Science, and citation indexes more generally, is still wanting~\cite{martin_google_2020}. Secondly, these proportions are not accounting for books, which are the primary means of publication in those fields of research. 

\begin{table}[ht]
	\caption{Number of citations equipped with identifiers (excluding identifiers matched via lookup), per type and compared with~\cite{halfaker_mansurov_redi_taraborelli_2018}. Note: a citation might be associated with two or more identifier types.}
	\centering
	\begin{tabular}{@{}l r r r p{0.07\linewidth}p{0.25\linewidth}p{0.32\linewidth}p{0.16\linewidth}@{}}
		\hline
		Id. & Our dataset & Previous work & Difference\\
		\hline
		DOI & 1,442,177 & 1,211,807 & 230,370\\
		ISBN & 2,160,818 & 1,740,812 & 420,006 \\
		PMC & 279,378 & 181,240 & 98,138\\
		PMID& 825,971 & 609,848 & 216,123\\
		ArXiv & 47,601  & 50,988 & -3,387\\
		Others & 308,268 & 0 & 308,268 \\
		\hline
		Total & 4,755,945  & 3,794,695 & 961,250 \\
		\hline
	\end{tabular}
	\label{tab:differencedataset}
\end{table}

\begin{table}[ht]
	\caption{Number of distinct DOI and ISBN identifiers across Wikipedia.}
	\centering
	\begin{tabular}{@{}l r r r p{0.15\linewidth}p{0.15\linewidth}p{0.15\linewidth}p{0.15\linewidth}@{}}
		\hline
		Category & Previously known & Newly found & Total\\
		\hline
		DOI & 1,018,542 & 153,879 & 1,157,571 \\
		ISBN & 901,639 & -- &  901,639 \\
		\hline
	\end{tabular}
	\label{tab:unique_identifiers}
\end{table}

\begin{figure}
	\centering
	\includegraphics[width=\linewidth]{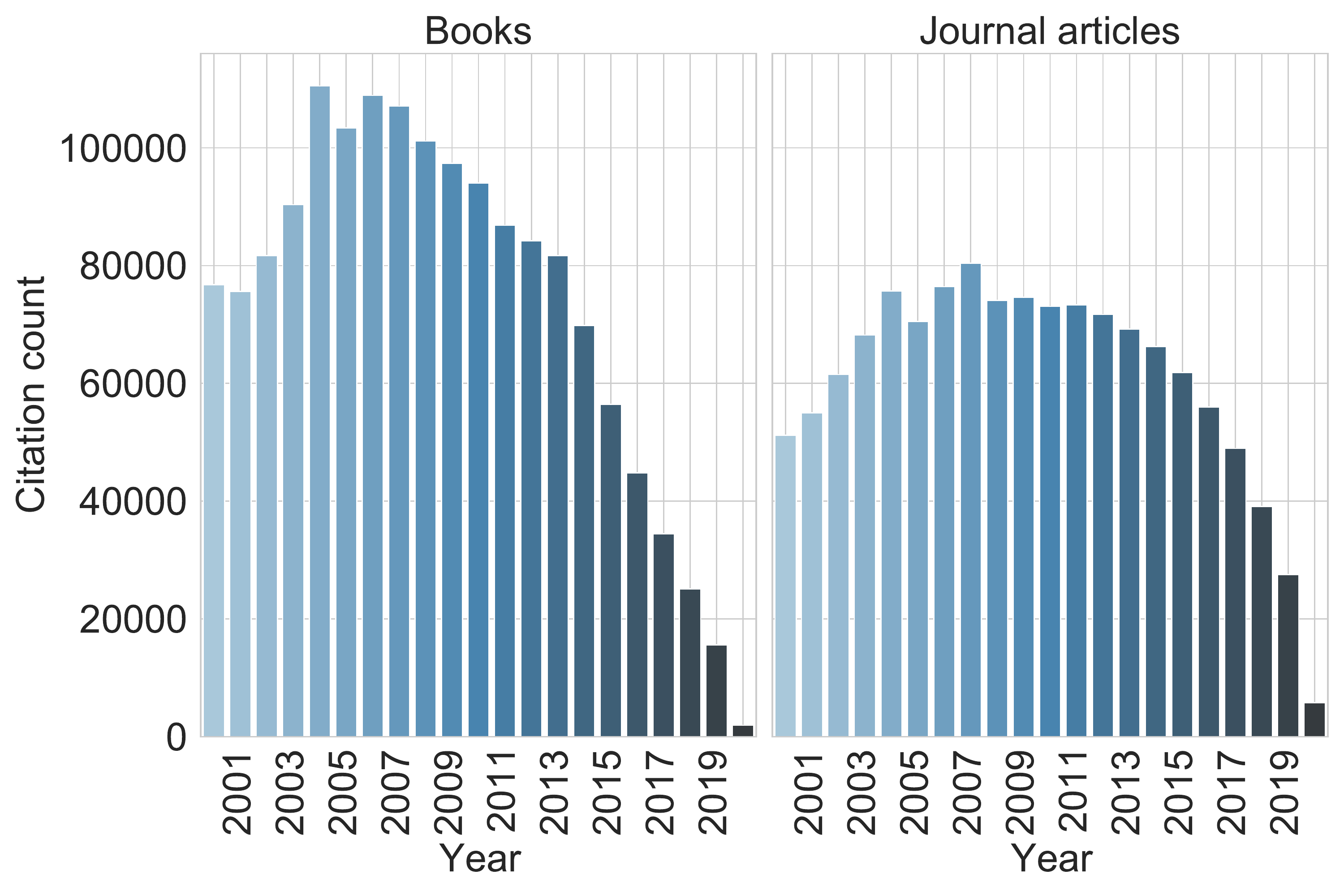}
	\caption{Publication years for \textit{journal articles} and \textit{books}, for the period 2000-2020.}
	\label{fig:top_publication_years}
\end{figure}

\begin{figure}
	\centering
	\includegraphics[width=\linewidth]{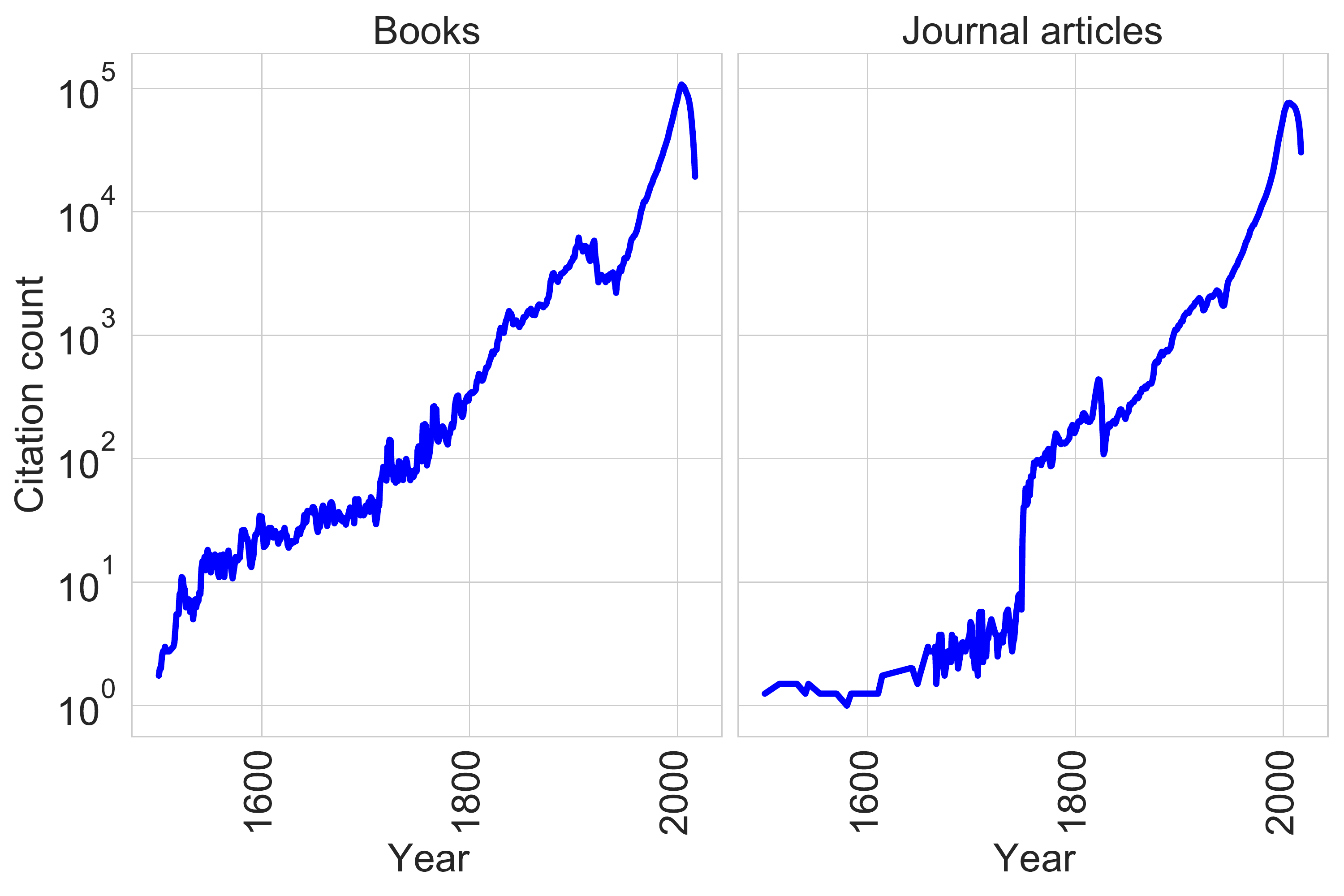}
	\caption{Number of citations per source publication year (1500-2020). A smoothing average using a window of 4 years is applied.}
	\label{fig:publication_for_all_years}
\end{figure}

We show in Figure \ref{fig:top_publication_years} the number of citations to books and journal articles published over the time period 2000 to 2020. This figure highlights how books appear to take longer to get cited in Wikipedia after publication. A similar plot, but considering a much wider publication time span (1500-2020) is given in Figure~\ref{fig:publication_for_all_years}. Most published material in Wikipedia dates from the 1800 onward. We note that a total of 89,098 journal article citations and 193,336 book citations do not contain a publication year.

Out of all the 28 template keys including the citation, most are not complete. For example, identifiers are present only in 13.42\% of citations whereas URLs are present in 85.25\% of citations. This implies that many citations refer to Web contents.

\begin{figure}
	\centering
	\includegraphics[width=\linewidth]{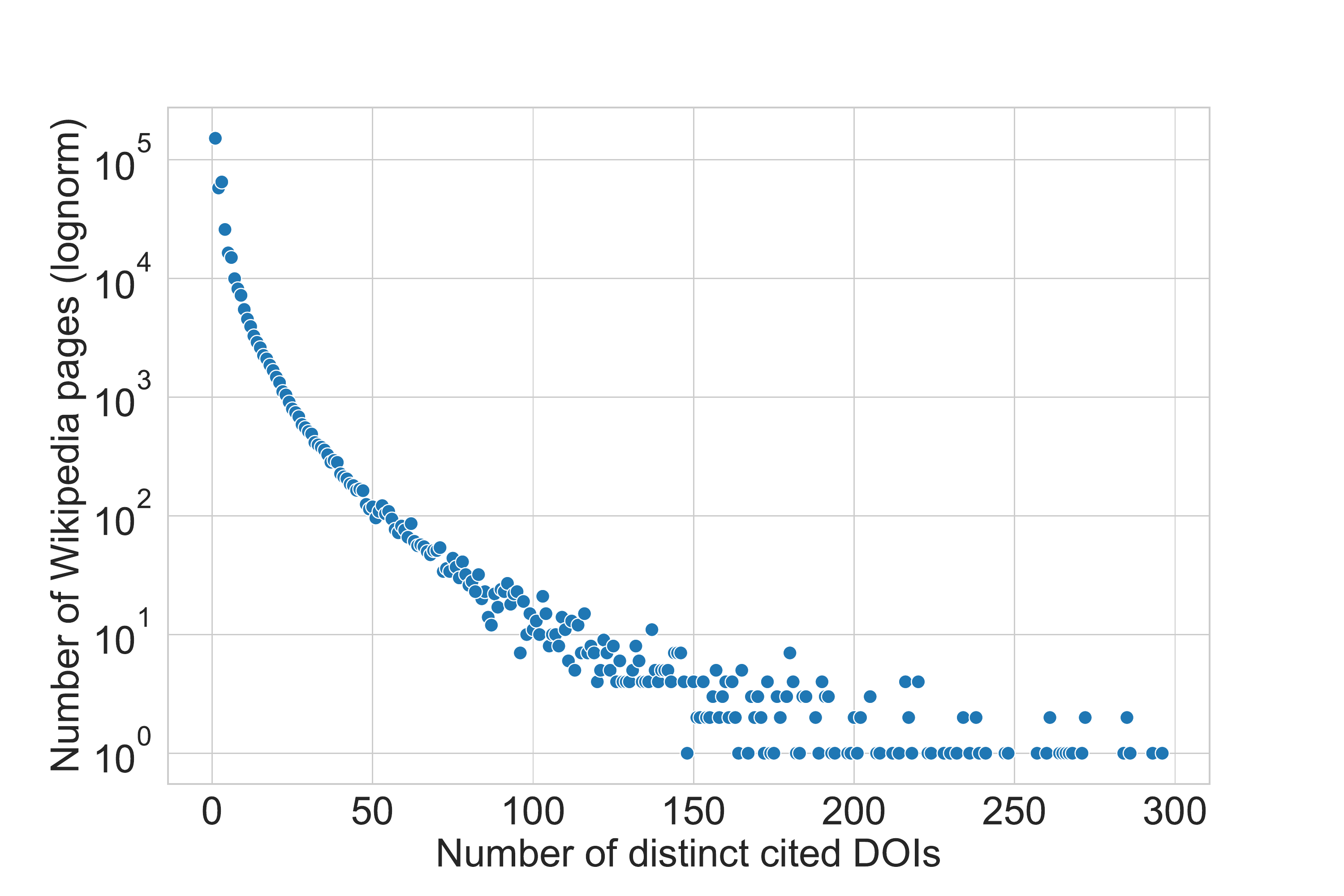}
	\caption{Number of distinct cited DOI per Wikipedia page.}
	\label{fig:hist_distinct_dois}
\end{figure}

Out of 6,069,685 pages on Wikipedia, 405,359 have at least one or more citations with a DOI, that is about 6.7\%; the proportion goes up to 12.84\% for pages with at least one ISBN instead. This higher percentage of pages with DOIs, when compared to previously reported values~\cite{maggio_wikipedia_2019}, is in large part due to our newly found identifiers from Crossref which allowed us to equip with DOIs citations coming from Wikipedia pages with no previous presence of DOIs. We eventually considered the distribution of distinct DOIs per Wikipedia page and it was found that most of the pages have few citations with DOI identifiers, as shown in Figure~\ref{fig:hist_distinct_dois}. The top journals are listed in Table \ref{tab:ten_most_cited_periodicals}, and contain well-known mega journals (Nature, Science, PNAS) or other reputed venues (Cell, JBC).

\begin{table}[ht]
	\caption{Most cited journals.}
	\centering
	\begin{tabular}{@{}l r p{0.7\linewidth}p{0.15\linewidth}@{}}
		\hline
		Journal Name & Citations\\
		\hline
Nature                                                                         & 36,136 \\
Science                                                                        & 26,448 \\
Journal of Biological Chemistry                                                                 & 22,401 \\
PNAS & 21,347 \\
The IUCN Red List of Threatened Species                                        & 10,082 \\
Cell                                                                           & 9329  \\
Zootaxa                                                                        & 8013  \\
Genome Research                                                                   & 6994  \\

		\hline
	\end{tabular}
	\label{tab:ten_most_cited_periodicals}
\end{table}

\subsection{Limitations and research opportunities}

\noindent The \texttt{Wikipedia Citations} dataset can be useful for research and applications in a variety of contexts. We suggest a few here, and also frame the limitations of our contribution as opportunities for future work.

\subsubsection{Map of Wikipedia sources} 
What seems to us a low-hanging fruit is creating a map of Wikipedia sources, following the science mapping and visualization methodologies~\cite{shiffrin_mapping_2004,borner_atlas_2010,chen_science_2017}. Such work would allow to comprehensively answer the question of what is cited from Wikipedia, from which Wikipedia articles, and how knowledge is reported and negotiated in Wikipedia. Importantly, such mapping should consider disciplinary differences in citations from Wikipedia, as well as books (5.3M citations by our estimates) and non-scientific sources such as news outlets and other online media (21.5M citations), which make up for the largest share of Wikipedia citations. Answering these questions is critical to inform the community work on improving Wikipedia by finding and filling knowledge gaps and biases, all the same guaranteeing the quality and diversity of the sources Wikipedia relies upon~\cite{mesgari_sum_2015,damato_provenance_2017,hube_bias_2017,piscopo_what_2019,wang_assessing_2020}.

\subsubsection{Citation reconciliation and recommendation} 
Link prediction in general, and citation recommendation in particular, have been explored for Wikipedia since some time~\cite{fetahu_finding_2016,paranjape_improving_2016,wulczyn_growing_2016}. Recent work has also focused on finding Wikipedia statements where a citation to a source might be needed~\cite{redi_citation_2019}. Our dataset can further inform these efforts, in particular easing and fostering work on the recommendation of scientific literature to Wikipedia editors. The proposed citation classifier could also be re-used for citation detection and reconciliation in a variety of contexts.

\subsubsection{Citations as features} 
Citations from Wikipedia can be used as `features' in a variety of contexts. They have already been considered as altmetrics for research impact~\cite{sugimoto_scholarly_2017}, while they can also be used as features for machine learning applications such as those focused on improving knowledge graphs, starting with Wikidata~\cite{farda-sarbas_wikidata_2019}. It is our hope that more detail and novel use cases will also lead to a gradual improvement of the first version of the dataset which we release here.

\subsubsection{Limitations}

We highlight a set of limitations which constitute possible directions for future work. First of all, the focus on English Wikipedia can and should be rapidly overcome to include all languages in Wikipedia. Our approach can be adapted to other languages, provided that external resources (e.g., language models and lookup APIs) are available for them. Secondly, the dataset currently does not account for the edit history of every citation from Wikipedia, while this would allow to study knowledge production and negotiation over time: adding `citation versioning' would be important in this respect, as demonstrated by recent work~\cite{zagovora_i_2020}. Thirdly, citations are used for a purpose, in a context; an extension of the \texttt{Wikipedia Citations} dataset could include all the citation statements as well, in order to allow researchers to study the fine-grained purpose of citations. Furthermore, the classification of scientific publications which we use is limited. ISBNs, in particular, can refer to monographs, book series, book chapters and edited books, which possess varying citation characteristics. Future work should go in extending the classification system to operate at such finer-grain level. Lastly, the querying and accessibility of the dataset is limited by its size; more work is needed in order to make Wikipedia's contents better structured and easier to query~\cite{aspert_graph-structured_2019}.

\section{Conclusion}
\noindent We publish the \texttt{Wikipedia Citations} dataset, consisting of a total of 29.276M citations extracted from 6.069M articles from English Wikipedia. Citations are equipped with persistent identifiers such as DOIs and ISBNs whenever possible. Specifically, we extracted 3.928M citations with identifiers --- including DOI, PMC, PMID, and ISBN --- from Wikipedia itself, and further equipped an extra 260,752 citations with DOIs from Crossref. In so doing, we were able to raise the number of Wikipedia pages citing at least one scientific article equipped with a DOI from less than 5\% to more than 6.7\% (which corresponds to an additional 164,830 pages) and found that Wikipedia is citing just 2\% of the scientific articles indexed in the Web of Science. We also release all our code to extend upon our work and update the dataset in the future. Our work contributes to ongoing efforts~\cite{halfaker_mansurov_redi_taraborelli_2018,zagovora_i_2020} by expanding the coverage of Wikipedia citations equipped with identifiers, distinguishing between academic and non-academic sources, and by releasing a codebase to keep results up-to-date.

We highlighted a set of possible uses of our dataset, from mapping the sources Wikipedia relies on, to recommending citations and using citation data as features. The limitations of our contribution also constitute avenues for future work. We ultimately believe that \texttt{Wikipedia Citations} should be made available as data via open infrastructures, e.g., WikiCite\footnote{\url{https://meta.wikimedia.org/wiki/WikiCite} [accessed 2020-11-12].} and OpenCitations\footnote{\url{https://opencitations.net} [accessed 2020-11-12].}. We consider our work a step in this direction. It is therefore our hope that this contribution will start a collaborative effort by the community to study, use, maintain and expand work on citations from Wikipedia.

\section*{Data availability}

The dataset is made available on Zenodo~\cite{dataset} and the accompanying repository contains all code and further documentation to replicate our results: \url{https://github.com/Harshdeep1996/cite-classifications-wiki/releases/tag/0.2}.

\section*{Acknowledgements}

The authors would like to thank Tiziano Piccardi (EPFL), Miriam Redi (Wikimedia Foundation) and Dario Taraborelli (Chan Zuckerberg Initiative) for their helpful advice. The authors also thank the Centre for Science and Technology Studies (CWTS), Leiden University, for granting access to the Web of Science.

\bibliographystyle{plainurl}
\bibliography{mybib}

\section*{Appendix}

\begin{table}[ht]
	\caption{Different citation templates can be used to refer to the same source.}
	\centering
	\begin{tabular}{p{0.25\linewidth}p{0.65\linewidth}}
		\hline
		Index & Extracted citation template\\
		\hline
		Citation 1 & \{\{citation$\vert$\textbf{author}=John Smith$\vert$ access\-date=February  17, 2006\}\}\\
		Citation 2 & \{\{citation$\vert$\textbf{creator}=John Smith$\vert$ access\-date=September 15, 2006\}\}\\
		\hline
	\end{tabular}
	\label{tab:differentkey}
\end{table}

\begin{table}[ht]
	\caption{Citations are mapped to have the same keys.}
	\centering
	\begin{tabular}{p{0.25\linewidth}p{0.65\linewidth}}
		\hline
		Index & Uniform citation template\\
		\hline
		Citation 1 & \{`author': `John Smith', `type': `citation', `accessdate': `February 17, 2006'\}\\
		Citation 2 & \{`author': `John Smith', `type': `citation', `accessdate': `September 15, 2006'\}\\
		\hline
	\end{tabular}
	\label{tab:differentdict__}
\end{table}

\begin{table}[ht]
	\caption{Example of newly-classified citations.}
	\centering
	\begin{tabular}{p{0.25\linewidth}p{0.7\linewidth}}
		\hline
		Label & Citation\\
		\hline
		Journal article & \{`title': `What is Asia?', `author': `Philip Bowring'\}\\
		Journal article & \{`title': `Right Ventricular Failure', `journal': `e-Journal of Cardiology Practice'\}\\
		\hline
		Book & \{`title': `Histories of Anthropology Annual, Vol. I', `author': `HS Lewis'\}\\
		Book & \{`title': `The Art of the Sale', `publisher': `The Penguin Press'\}\\
		\hline
		Web content & \{`title': `Barry White - Chart history (Hot R\&B\/ Hip-Hop Songs) Billboard', `page\_title':  Let the Music Play (Barry White Album)'\}\\
		Web content & \{`title': `Sunday Final Ratings: Oscars Adjusted Up', `work': `TVbytheNumbers'\}\\
		\hline
	\end{tabular}
	\label{tab:differentdict}
\end{table}

\begin{table}[ht]
	\caption{Presence of identifiers per citation for the 3.92 million citations with identifiers (with 0 = False and 1 = True). These counts sum up to 3,620,124, with an additional 308,268 citations associated with other identifiers such as OCLC, ISSN. The total adds up to 3,928,392 citations with identifiers.}
	\centering
	\begin{tabular}{@{}r r r r r r r p{0.05\linewidth}p{0.1\linewidth}p{0.1\linewidth}p{0.1\linewidth}p{0.1\linewidth}p{0.1\linewidth}p{0.1\linewidth}@{}}
		\hline
		With DOI & With ISBN & With PMC & With PMID & With ARXIV & Total\\
		\hline
		0 & 0  & 0 & 0 & 1 & 4,447 \\
		0 & 0  & 0 & 1 & 0 & 41,417 \\
		0 & 0  & 0 & 1 & 1 & 7 \\
		0 & 0  & 1 & 0 & 0 & 829 \\
		0 & 0  & 1 & 1 & 0 & 11,261 \\
		0 & 0  & 1 & 1 & 1 & 5 \\
		0 & 1  & 0 & 0 & 0 & 2,119,545 \\
		0 & 1  & 0 & 0 & 1 & 192 \\
		0 & 1  & 0 & 1 & 0 & 223 \\
		0 & 1  & 1 & 0 & 0 & 13 \\
		0 & 1  & 1 & 1 & 0 & 8 \\
		1 & 0  & 0 & 0 & 0 & 592,557 \\
		1 & 0  & 0 & 0 & 1 & 35,824 \\
		1 & 0  & 0 & 1 & 0 & 501,176 \\
		1 & 0  & 0 & 1 & 1  & 5,101 \\
		1 & 0  & 1 & 0 & 0 & 4,241 \\
		1 & 0  & 1 & 0 & 1 & 3 \\
		1 & 0  & 1 & 1 & 0 & 261,173 \\
		1 & 0  & 1 & 1 & 1 & 1,265 \\
		1 & 1  & 0 & 0 & 0 & 35,706 \\
		1 & 1  & 0 & 0 & 1 & 756 \\
		1 & 1  & 0 & 1 & 0 & 3,794 \\
		1 & 1  & 0 & 1 & 1 & 1 \\
		1 & 1  & 1 & 0 & 0 & 40 \\
		1 & 1  & 1 & 1 & 0 & 540 \\
		\hline
	\end{tabular}
	\label{tab:identifier_per_citation}
\end{table}

\begin{table}[ht]
	\caption{Web of Science 30 most-represented subject categories, by number of articles cited from Wikipedia. We consider the first subject category of each article, and discard the rest. The total number of articles which we could match in the Web of Science using DOIs is 710,913. The top-30 subject categories make up almost 60\% of them.}\label{tab:wos_sc}
	\centering
\begin{tabular}{lr}
\toprule
Web of Science subject category &  Number of articles \\
\midrule
Biochemistry \& Molecular Biology            &             81,556 \\
Multidisciplinary Sciences                  &             51,368 \\
Astronomy \& Astrophysics                    &             18,658 \\
Medicine, General \& Internal                &             17,134 \\
Neurosciences                               &             16,061 \\
Cell Biology                                &             14,411 \\
Genetics \& Heredity                         &             13,922 \\
Chemistry, Multidisciplinary                &             13,754 \\
Microbiology                                &             13,661 \\
Oncology                                    &             13,606 \\
Plant Sciences                              &             13,050 \\
Clinical Neurology                          &             13,017 \\
Immunology                                  &             11,231 \\
Biotechnology \& Applied Microbiology        &             10,980 \\
Pharmacology \& Pharmacy                     &             10,351 \\
Ecology                                     &             10,308 \\
Zoology                                     &             10,081 \\
Biology                                     &              8979 \\
Endocrinology \& Metabolism                  &              8444 \\
Geosciences, Multidisciplinary              &              7653 \\
Public, Environmental \& Occupational Health &              7348 \\
Biochemical Research Methods                &              7337 \\
Economics                                   &              7211 \\
Physics, Multidisciplinary                  &              7042 \\
Paleontology                                &              6942 \\
Behavioral Sciences                         &              6591 \\
Environmental Sciences                      &              6316 \\
Chemistry, Physical                         &              6292 \\
History                                     &              6174 \\
Cardiac \& Cardiovascular Systems            &              5875 \\
\midrule
Total & 425,353 (60\%)\\
\bottomrule
\end{tabular}
\end{table}

\end{document}